# Xerogel Protein Substrates for Far Infrared Studies

J.-Y. Chen,[a] W. Cox,[a,b] E. Tehan,[b] F.V. Bright,[b] J. Cerne[a] and A. G. Markelz[a]
a)Physics Department, University at Buffalo
b) Chemsitry Department, University at Buffalo

## Abstract

A number of researchers are attempting to develop a biosensor based on terahertz sensing. Key to the realization of any biosensor system is the access of the target molecules to the sensor. In the case of using terahertz spectroscopy, there are limitations due to high attenuation by water and sensitivity levels. In order to have a significant response at THz frequencies generally requires a minimum pathlength of 100 μm. Solid thin films of this thickness will have limited interaction with target molecules. We propose to use xerogels as biomolecular probe substrates. We have characterized the THz transmission of xerogel substrates as a function of hydration and protein content. These measurements demonstrate that xerogels are excellent candidates for biomolecular probe substrates to realize THz biosensors.

## I. Introduction

Biomolecular fingerprinting has long been an optical challenge. While some biomolecules may have a distinct optical absorption, circular dichroism feature or fluorescence resonance, such features are not general for all biomolecules. Thus a generalized biosensor using a single optical property to identify the presence of a biomolecule can not be realized. As a result many biosensor schemes are based on bioaffinity. Here the binding of a target biomolecule to a highly selective complement, or probe, is sensed by monitoring a physical characteristic of the probe. Often this physical characteristic is one engineered into the probe, such as a fluorescent tag. A more ideal determination of binding would be to use an intrinsic physical characteristic that must change with binding. We have proposed using the terahertz dielectric properties of the probe to determine if binding has occurred. The vibrational modes of biomolecules associated with collective movement of the tertiary structure lay in the terahertz frequency range. We have shown previously that the THz absorbance of many biomolecular samples closely resemble the calculate density of normal modes.[1] The frequency and strength of the absorbance will be affected by target binding due to induced conformational change, induced dipole change or simply due to mass change. Such a method is free from tagging of probe molecules.

We have demonstrated that THz absorbance and dielectric response is dependent on ligand binding for thin films.[2] In these studies, solutions of probe molecules and probe molecules complexed with a target were used to form thin films. In real world applications this type of sample preparation would be impractical. Instead one would wish to monitor the THz transmission of immobilized probe molecules on a substrate and the presence of the target bound to the probe would be determined by the change in the THz transmission. Key to the realization of any biosensor system is the access of the target molecules to the sensor. Thin films of probe molecules would provide limited target access in that the binding would be mainly confined to the surface of the film. It would be preferable to have a substrate in which both the probe is encapsulated, and the target has ready access to the probe. We investigate the application of probe molecules embedded in xerogels as a method to increase target access and optical density. Xerogels are high porosity nanoporous silica films. Biomolecular inclusion in xerogels has been demonstrated and in most cases the biomolecule behaves as in a dilute solution.[3] The pore size for these xerogels is determined by the embedded biomolecule and the silica network allows for sufficient access to the embedded biomolecule so that reactions such as change of pH occur as if in solution. The use of xerogels not only eliminates the concern for target access it also allows thick samples to rapidly reach their equilibrium hydration at a given atmospheric relative humidity. Xerogels have not before been characterized in the THz frequency range. A key question is how the transparency is affected with relative humidity. For example, fused quartz windows are nearly opaque in the THz range due to water within the glass. An additional question is the absorbance of proteins measured in the xerogels versus that measured either in thin films or pellets. Here we report our measurements of the absorbance and index measured for xerogels as a function of relative humidity and how the transmission of the xerogels are affected by biomolecular inclusion, using hen egg white lysozyme (HEWL) as our model biomolecule. The data shows that the xerogels are sufficiently transparent in the THz range for use as substrates. The biomolecular inclusion does not affect the transmission of the xerogel in a simple additive way, but rather changes the nature of the dielectric response of the system as a whole. We suggest that this change is due to a change in the way water can bind to the xerogel in the presence of the biomolecule.

## II. Materials and Methods

Tetramethyl ortho-silicate (TMOS), and HEWL were purchased without further purification (Sigma 34,143-6 and Sigma 6876 respectively). To form the sol-gel solution, 648.9uL of 0.04N (0.04M) HCl was added to 17.96 g of TMOS and 3.488mL of $H_2O$. The solution was sonicated for 20 minutes. After sonication, 2mL of the sol was mixed with 2 ml of buffer (protein solution) to form the reference (sample) solution. These precursor solutions were then additionally sonicated 20 minutes before pouring into polystyrene cuvettes. The reference buffer was a pH 7 Tris buffer, and the protein solution was a 40 mg lysozyme to 1 ml Tris buffer pH 7 solution. Solidification began within 30 seconds of adding the solution.[3] Solvent was allowed to evaporate over time. Samples were cut with a diamond saw and plates were formed with thicknesses of 610-660 μm.

A standard THz time domain spectroscopy (TTDS) system is used to monitor the change in absorbance and index with oxidation state.[4] The THz is generated using a hertzian dipole antenna and detected using electro optic detection or antenna detection using a photoconductive switch on ion implanted silicon on sapphire.[5,6] Samples are mounted in a humidity controlled cell.[7] Hydrations studied here were dry (< 5% r.h.), 33 % r. h., 53 % r.h., and hydrated, (80 % r.h.). The hydration cell is flushed with either dry gas or gas hydrated by flowing over a saturated salt solution. For oxidized samples the flow gas is oxygen, for the reduced samples the flow gas is air. The typical time for flushing before data taking was >2 h. By performing the measurements as a function of hydration we can get some idea of the role water plays in the change in the THz response. All measurements were performed at room temperature. TTDS, which measures the transmitted *field*, allows us to both access the real and imaginary part of the dielectric response in a single measurement. Thus we can both see how the index and the absorbance change as a function of frequency. The field transmission is determined by the measured transmitted field for the sample and reference through:

$$t = \frac{E_{sample}(\omega)e^{i\phi_{sample}(\omega)}}{E_{ref}(\omega)e^{i\phi_{ref}(\omega)}} = |t|e^{i\phi_t(\omega)} \quad (1)$$

Where $\phi_t(\omega)$ is the phase. The absorbance and index are then given by:

$$A(\nu) = -2\log(|t|), \quad n(\nu) = 1 + \frac{\phi_t(\nu)c}{2\pi\nu d} \quad (2)$$

where $\nu$ is the frequency in terahertz, c the speed of light and d the sample thickness.

### III. Results and Discussion

In Figure 1 we show the absorbance derived from the measured transmission for the 660 μm thick xerogel for relative humidities of < 5% to 80 %, using Eq. 2. The large absorbance increases with frequency similar to the dielectric relaxation type response seen for many glasses. The periodic peaks seen in the data are due to multiple reflection (etalon) effects in the xerogel plate. This type of absorbance may arise from localized collective vibrations within the glass or from water that is still trapped within the glass.

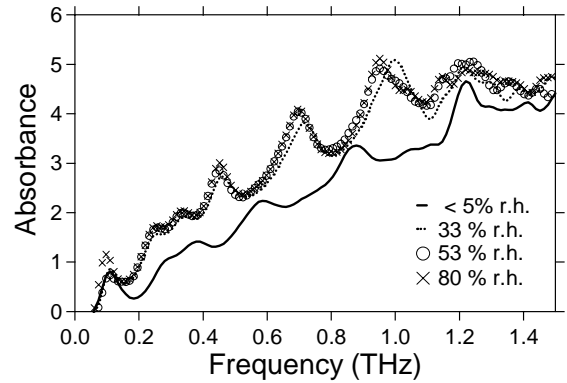

**Figure 1. THz absorbance of a 660 μm xerogel plate as a function of relative humidity.**

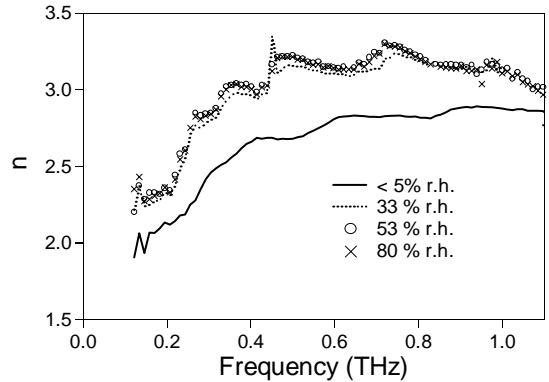

**Figure 2. Index measured for 660 μm thick xerogel.**

As the humidity increases it is expected that the absorbed water will both contribute to an increase in absorbance and index. As seen in Fig. 1 the absorbance does increase somewhat for 33% r.h. but at higher humidities there is no further increase. This result is somewhat surprising and suggests that the additional water absorbed does not contribute to the THz absorbance. While the absorbance is high, the transparency of the xerogel is sufficient for THz measurements of embedded proteins.

In Figure 2 we show the index of the neat xerogel for different humidities. As seen the dielectric constant increases with hydration and this increase has saturated by 33 % r.h. in agreement with the absorbance measurements. The shift in the dielectric constant is in good agreement with the shift in etalon fringes seen in the absorbance data. The net increase in dielectric constant with hydration corresponds to an effective dielectric constant determined by the combination of the adsorbed water and the silica glass.

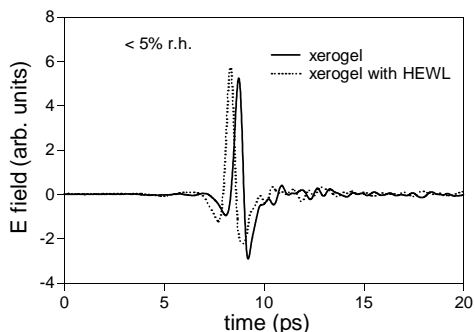

Figure 3. Transmitted waveforms for xerogel and xerogel with HEWL at < 5% r.h.

In Figure 3 we show the waveforms for the transmitted THz pulse for the neat xerogel and the xerogel with lysozyme at < 5 % r.h.. The thickness of the lysozyme sample is 610 μm. As seen the pulse transmitted through the lysozyme sample has an earlier arrival time and higher peak value than the neat xerogel! To first order this result is somewhat surprising, since one might expect that the addition of the biomolecules would increase absorbance and index given that a) the biomolecules filling the pores of the xerogel are a) more highly absorbing than air and b) have a higher index than air. While the time delay and peak difference for the low hydration samples can be accounted for in part by the thickness difference between the two samples. However the difference increases in a striking way at higher hydrations as shown in Figure 4 for 80 % r.h. The difference in transmitted waveforms shows that the index and absorbance of the lysozyme sample is smaller than the neat xerogel. We suggest several possible sources for the observed results. The lysozyme binding within the pores may be reducing the surface available for molecular water binding. In addition the larger pore size of the lysozyme embedded xerogel could reduce the net silica surface area available for water binding. This change in the nature of the material with the embedding of the biomolecule prevents us from making absolute THz dielectric characterization of biomolecules using xerogel substrates, however the fact that the molecules within the xerogel can be fully hydrated and targets can access probes means that using xerogel substrates to measure relative change in THz absorbance with ligand binding is viable.

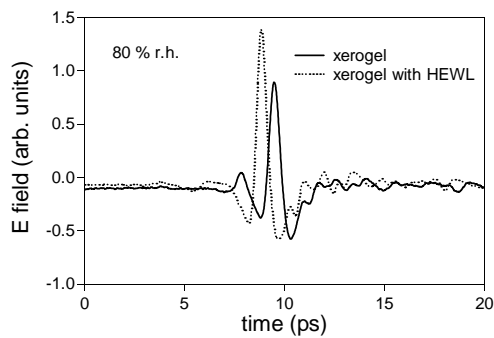

Figure 4. Transmitted THz waveforms for xerogel and xerogel with HEWL at 80 % r.h.

## IV.   Conclusions

We have performed the first dielectric characterization of xerogels in the terahertz frequency range. The xerogels have an absorption coefficient of 174 cm$^{-1}$ and index ~ 3.16 at 33 % r.h. and 1 THz which is sufficiently low to use xerogels as substrates for THz measurements of biomolecular samples. Inclusion of biomolecules into the xerogel does not result in a simple additive response, but rather the biomolecules within the xerogel define a new composite material with a distinct THz dielectric response. The relative transparency of the composite material suggests that xerogels are viable substrates for realizing THz biosensor systems.

**Acknowledgements**
We gratefully acknowledge the support of this work by the Army Research Office, grant DAAD19-02-1-0271.